# Application of the Mössbauer spectroscopy to study harmonically modulated electronic structures: case study of charge- and spin-density waves in Cr and its alloys

*S.M. Dubiel*[*]

*AGH University of Science and Technology, Faculty of Physics and Applied Computer Science, PL-30-059 Kraków, Poland*

**Abstract**

Relevance of the Mössbauer spectroscopy in the study of harmonically modulated electronic structures i.e. spin-density waves (SDWs) and charge-density waves (CDWs) is presented and discussed. First, the effect of various parameters pertinent to the SDWs and CDWs is outlined on simulated [119]Sn spectra and distributions of the hyperfine field and the isomer shift. Next, various examples of the [119]Sn spectra measured on single-crystals and polycrystalline samples of Cr and Cr-V are reviewed.

[*]Stanislaw.Dubiel@fis.agh.edu.pl

## 1. Introduction

There are crystalline systems in which the electronic structure is harmonically modulated. The modulation of charge is known as charge-density waves (CDWs), and the one of spin as spin-density waves (SDWs). If both quantities are modulated one speaks about a co-existence of the CDWs and SDWs. The CDWs were found to exist in quasi-1D linear chain compounds like $TaS_3$ and $NbSe_3$, *2D* layered transition-metal dichalcogenides such as $TaS_2$, $VS_2$, or $NbSe_2$, *3D* metals like ω-Zr and Cr or compounds like $Mn_3Si$ and $UCu_2Si_2$ [1]. The best known system with SDWs is metallic Cr. Its SDWs which originate from *s*- and *d*-like electrons and show a variety of interesting properties were studied with different experimental techniques [2]. The most fundamental is their relationship to a density of electrons at the Fermi surface (FS). Between the Néel temperature of ~313 K and the so-called spin-flip temperature, $T_{SF} \approx 123$ K, the SDWs in chromium are transversely polarized i.e. the wave vector, $\underline{q}$, is perpendicular to the polarization vector, $\underline{p}$. Below $T_{SF}$ they are longitudinally polarized. One of the basic parameters pertinent to the harmonically structures is periodicity, $\Lambda$. The modulation is commensurate with the lattice if $\Lambda = n \cdot a$, where *a* is the lattice constant and *n* is an integer, the modulation is incommensurate if $\Lambda \neq n \cdot a$. The commensurability or incommensurability can be also expressed in terms of the wave vector which for the commensurate structures fulfils the equation $\underline{q} = 2\pi/a$. For the incommensurate structures, like the one in chromium, $\underline{q} \neq 2\pi/a$. The latter feature can be measured by a parameter $\delta$, such that $\underline{q} = 2\pi(1-\delta)/a$. Thus, the periodicity can be also expressed as $\Lambda^* = a/(1-\delta)$. $\Lambda^* = a$ for the commensurate structures and $\Lambda^* > a$ for incommensurate ones. This definition of the periodicity is quite unfortunate, especially for the commensurate structures, as it leads to terming "commensurate" the usual antiferromagnetic structure which has nothing to do with the harmonic modulation. The periodicity of incommensurate structures can be also measured using the following definition: $\Lambda_a = a/\delta$. In the case of chromium SDWs are incommensurate and their periodicity $\Lambda_a$ varies continuously between ~20 *a* at 4 K and ~28 *a* at RT [1].

Concerning the application of the Mössbauer spectroscopy (MS) to study the harmonically modulated electronic structures its relevance was recognized already in 1961 by Wertheim who made the first attempt to study the SDWs in Cr using the effect on [57]Fe nuclei which were introduced as probe atoms [3]. However, the trial was unsuccessful as the measured low temperature spectrum did not show any magnetic splitting. Instead, it had the form of a slightly broadened single line. The first successful application of MS was that by Street and Window who used the effect on [119]Sn nuclei introduced into Cr matrix as probe atoms [4-6]. The main difference in the two experiments lies in the fact that [57]Fe atoms are magnetic whereas [119]Sn ones are not. Theoretical calculations predict that magnetic impurities strongly interact with the SDWs causing their pinning i.e. the amplitude of the SDWs measured on the magnetic probe atom is zero or very small. This obviously was the reason for the failure of the Wertheim's experiment. On the other hand, non-magnetic atoms do not interact with the SDWs i.e. they do not disturb them; consequently they can be used as suitable probe atoms. The experiments carried out by Street and Window gave evidence for that [4-6]. Since then several experimental and theoretical papers relevant to the issue were published [7-25]. Concerning the former, the following ones are worth mentioning: (1) observation of the spin-flip transition in Cr [7], (2) determination of the third-order harmonics of the SDWs in Cr [8], (3) determination of the effect of grain boundaries on the SDWs in Cr [14], (4) determination of the effect of vanadium on the SDWs in Cr [18], (5) study of the critical behavior of Cr around the Neel temperature [20], (6) observation of a huge spin-density enhancement in the pre surface zone of a single-crystal Cr [21-23,25]. In theoretical papers pertinent to the SDWs and CDWs studies depicted the effect of (a) spin-density parameters such as periodicity, amplitude and sign of higher-order harmonics on (a) the [119]Sn spectra and hyperfine field distributions (for SDWs) [9,15,16], (b) the [119]Sn spectra and charge-density distributions (for CDWs) [15], and (c) the [119]Sn spectra and electric field gradient harmonic modulation [19].

## 2. Selected results and their discussion

### 2.1. CDWs

Coulomb interaction between two hole surfaces of the FS is the source of the CDWs. The CDW's order parameter is proportional to the square of the SDW's order parameter [26], so the CDWs can be expressed in terms of a series of even harmonics:

$$CDW = I_o + \sum_{i=i}^{N} I_{2i} \sin(2i\alpha + \varphi) \qquad (1)$$

Where $I_o$ is the average charge-density and $I_{2i}$ is the amplitude of the 2*i*-th harmonics, $\alpha = \mathbf{Q} \cdot \mathbf{r}$ (Q is the wave and **r** is the position vector) and $\varphi$ the phase shift. The calculated value of $I_2$ relative to $I_o$, ranges between $10^{-3}$ and $10^{-2}$ [27,28] which agrees well both with the X-ray diffraction [29] and $^{119}$Sn Mössbauer [30] measurements.

**2.1.1. Simulated $^{119}$Sn Mössbauer spectra**

A Mössbauer spectrum can be simulated taking into account one periodicity, $\Lambda_a$. At first a number of higher-order harmonics and their amplitudes are selected and the resulting CDW is constructed. Next, $\Lambda_a$ is divided into *N* equal intervals as shown in Fig. 1a. A single-line sub spectrum spectrum having the Lorentzian shape, having 1 mm/s full width at half maximum and the isomer shift, *S,* proportional to the amplitude of the CDW in the corresponding interval is next constructed. The sum of all *N*-sub spectra gives the final spectrum corresponding to the chosen CDW. As seen in Fig. 1c, its shape significantly differs from the Lorentzian shape and it rather resamples a broadened doublet. The distribution of the isomer shift can be derived either from the CDW itself or from the spectrum as indicated in Fig. 1d.

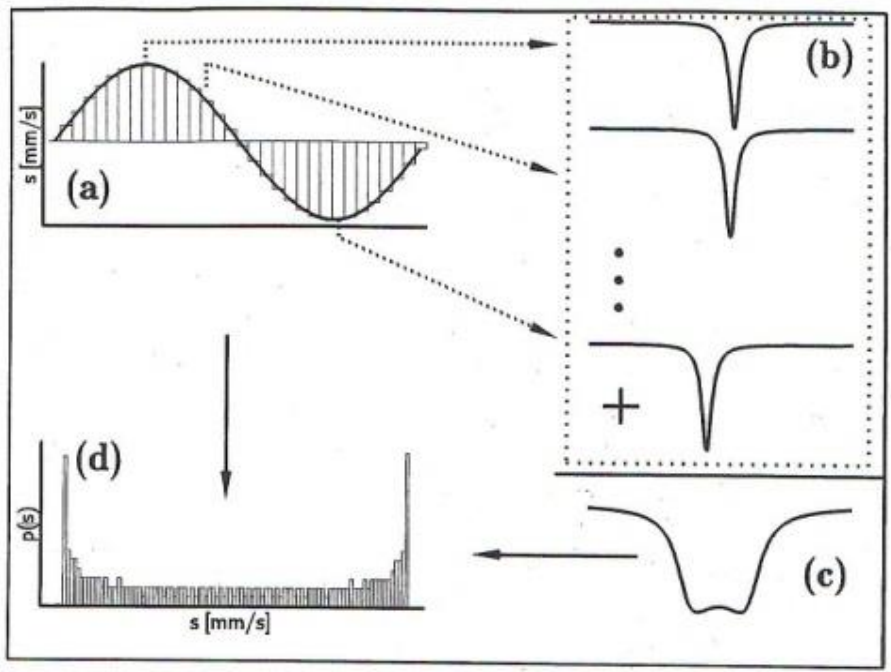

Fig. 1 Scheme showing the construction of a Mössbauer spectrum for the given shape of the CDW (one periodicity) displayed in (a). Presented in (b) are three of *N* sub spectra, each of which corresponds to one interval into which the CDW wave was divided. The overall spectrum is presented in (c), and the histogram of the isomer shift distribution can be seen in (d) [15].

**2.1.1.2. Commensurate CDWs, $\Lambda = n \cdot a$**

The shape of the spectra is sensitive both to the periodicity, *n*, as well as to the phase shift, $\varphi$ [15,24]. However, the difference between the spectra is large for small values of *n* (<10) while for *n>10* it becomes smaller and eventually does not depend either on *n* or $\varphi$.

**2.1.1.3. Incommensurate CDWs, $\Lambda \neq n \cdot a$**

In this case the spectra are sensitive to the amplitudes and signs of harmonics as well as to the phase shift. To illustrate these effects a set of spectra and underlying distributions of the isomer shift are shown in Fig. 2. It is evident that both the amplitude as well as phase shift can significantly affect the shape of the spectra.

**2.2. SDWs**

The Coulomb interaction between the electron and the hole surfaces of FS gives rise to the SDWs [31]. They can be expressed in terms of a series of odd harmonics:

$$SDW = \sum_{i=1}^{N} H_{2i-1} \sin\{(2i-1)\alpha + \varphi)\} \qquad (2)$$

In the following will be presented the effect of the periodicity (for the commensurate SDWs), amplitudes and signs of harmonics on the shape of the $^{119}$Sn Mössbauer spectra and underlying histograms of the spin-density distributions.

**2.2.1. Simulated spectra and spin-density distributions**

**2.2.2. Protocol of construction**

The hyperfine field has the inversion symmetry hence for a construction of a spectrum characteristic of the SDW it is enough to consider the half periodicity. It is divided into *N* equally spaced intervals, and for which of them a sub spectrum is constructed with the splitting proportional to the amplitude of the SDW in a given interval. The shape of the lines is Lorentzian, the line width at half maximum equal to 1 mm/s and the relative intensities of the lines within the sextet equal to 3:2:1. Examples of the sub spectra and the overall one obtained by summation of the all N sub spectra are shown in Fig. 3. A histogram of the hyperfine field distribution is presented as well. It can be derived numerically either from the spectrum itself or from the shape of the SDW. The histogram is known as the Overhauser profile and its analytical formula is as follows [9]:

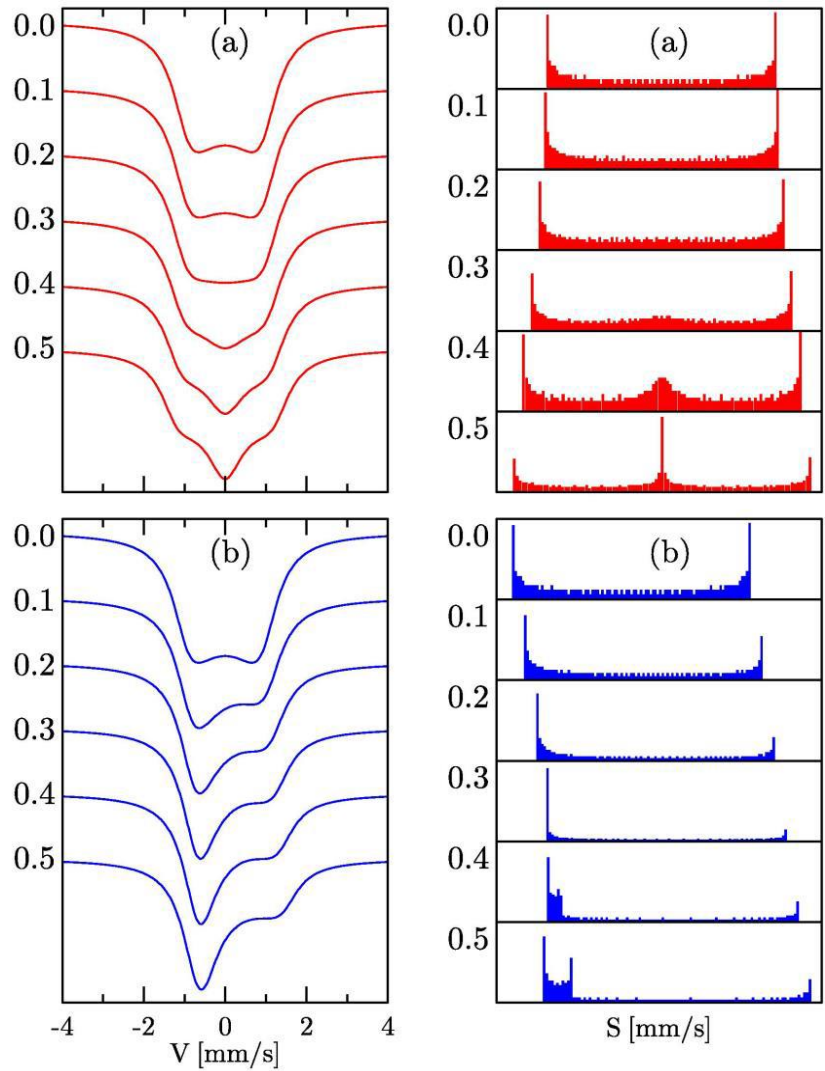

Fig. 2

Simulated $^{119}$Sn spectra and corresponding histograms of the isomer shift distribution for (a) CDW=0.5sin$\alpha$+$I_4$sin2$\alpha$, and (b) CDW=0.5sin2$\alpha$+$I_4$cos4$\alpha$. The figures labelling the spectra and the histograms stand for the value of $I_4$ [15].

$$p(H) = \frac{2}{\pi} \frac{1}{\sqrt{H_o^2 - H^2}} \qquad (3)$$

The most characteristic features of the *p(H)*, the probability of finding a *H* value in the range *[H, H+dH]*, is: (a) a cut-off at the most probable field, *H$_o$*, and a tail for *H<H$_o$*.

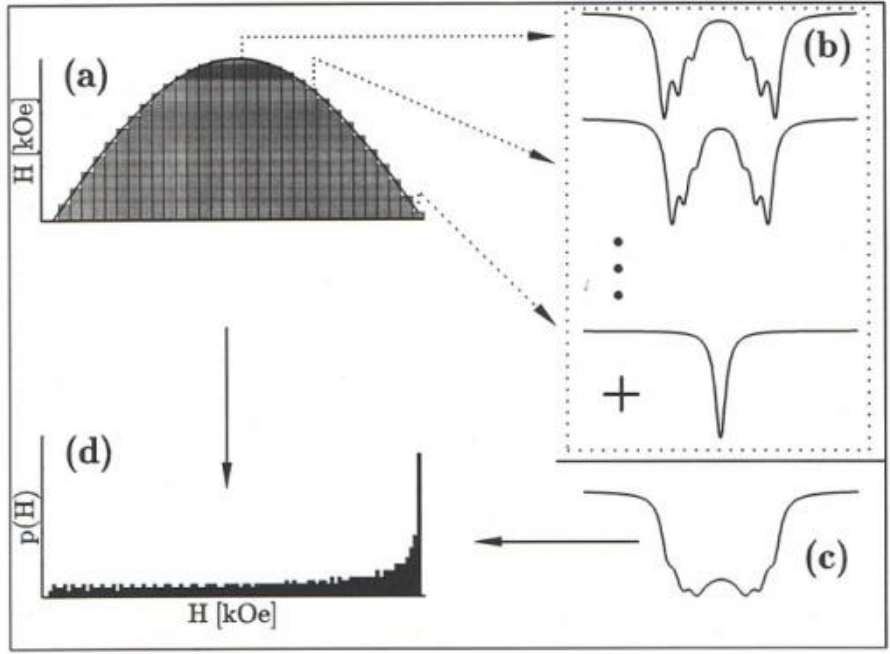

Fig. 3

Scheme showing the construction of a Mössbauer spectrum for the given shape of the SDW (half period) displayed in (a). Three particular sub spectra are presented in (b), the overall spectrum can be seen in (c), and the histogram of the hyperfine field distribution is visualized in (d) [13].

### 2.2.3. Commensurate SDWs, *Λ = n·a*

In this case the effect of the periodicity, *n*, and that of the phase shift, *φ*, can be investigated. Two cases are illustrated in Fig. 4 viz. (a) *φ*=0° (blue) and (b) *φ*=90° (red). It is clear that the actual shape of the spectrum depends on *n*. The dependence is especially strong for low values of *n*. Two further features are evident viz. (a) for *Λ=2n·a* the shape of the spectra is the same for the two, and (b) for higher *n*-values the difference between the two types of the spectra becomes smaller and the spectrum's shape approaches the one characteristic of the incommensurate SDWs (compare the spectrum for *n*=30 with the one shown in Fig. 5 for H$_1$=60 kOe).

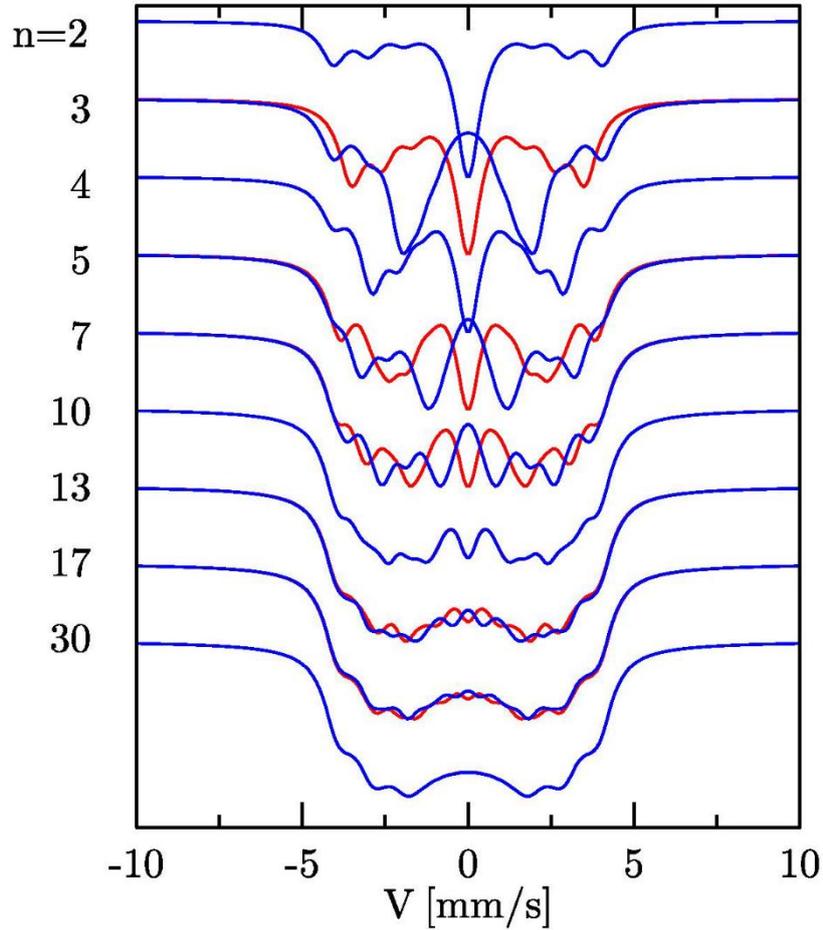

Fig. 4

Simulated $^{119}$Sn spectra for the commensurate SDW with the periodicity ranging between *n*=2 and *n*=30. Two cases are shown: (a) H=60·sinα (blue) and (b) H=60·cosα (red).

### 2.2.4. Incommensurate SDWs

#### 2.2.4.1. Fundamental harmonic, H$_1$

The purely sinusoidal SDW can be described as follows:

$$H = H_1 \sin \alpha \qquad (4)$$

In this case the effect of the amplitude, $H_1$, can be figured out. A set of the spectra obtained for $H_1$ ranging between 20 and 100 kOe is shown in Fig. 5. Noteworthy, the spectrum labelled with 60 is similar to the real spectrum measured at 295 K on a single-crystal sample of chromium [8] while the one labelled with 100 resembles the spectrum measured at 4.2 K on the same sample [20].

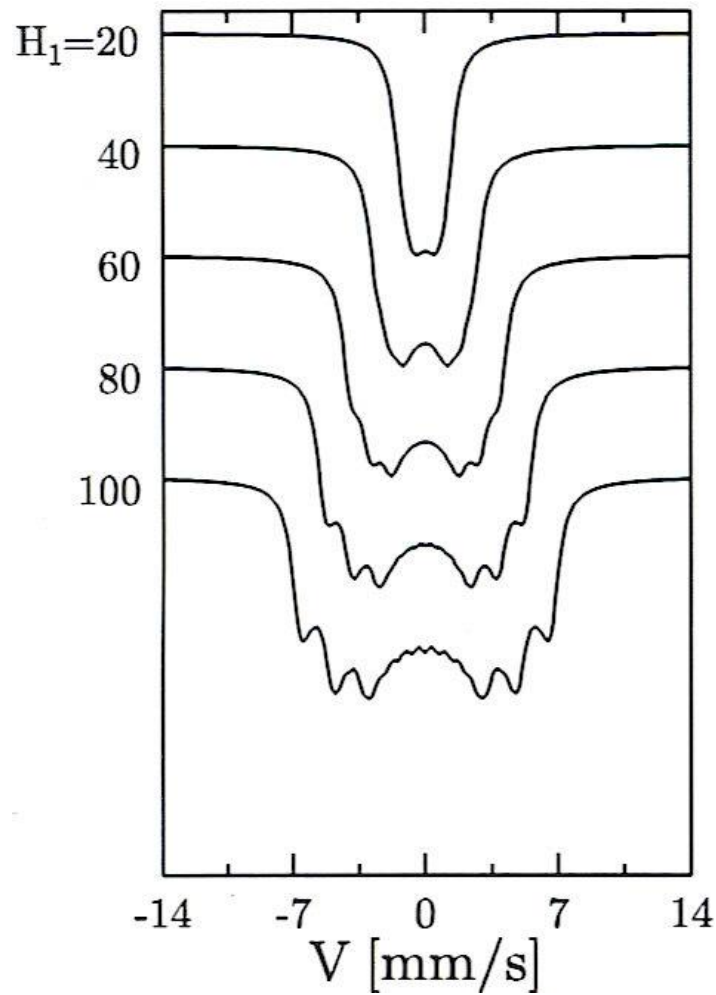

Fig. 5

$^{119}$Sn spectra simulated for the incommensurate SDW $H=H_1\sin\alpha$ with different values of $H_1$ in kOe.

**2.2.4.2. Third-order harmonic, $H_3$**

This harmonic is of particular interest because it is the second most important, and, it was revealed in chromium, the system in which the existence of the SDWs has been well evidenced with neutrons [31] and Mössbauer spectroscopy [8].

To illustrate the effect of $H_3$ and its sign $^{119}$Sn spectra were simulated for the following two SDWs:

$$H = H_1 \sin\alpha + H_3 \sin 3\alpha \qquad (5a)$$

$$H = H_1 \sin\alpha - H_3 \sin 3\alpha \qquad (5b)$$

The value of $H_1$ was kept constant at 60 kOe while $H_3$ was changed between 0 and 15 kOe. The output of the simulations is displayed in Fig. 6 (spectra), and in Fig. 7 (histograms).

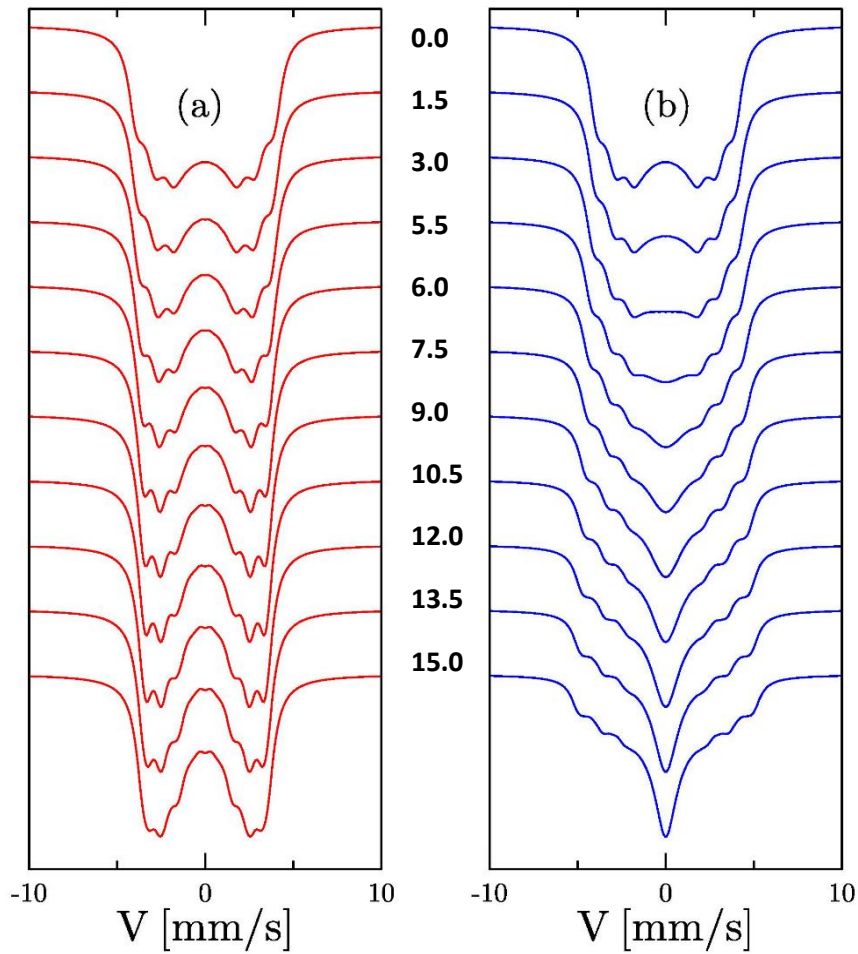

Fig. 6

$^{119}$Sn spectra simulated for different values of the third-order harmonics, $H_3$, ranging between 0.0 and 15 kOe. The spectra shown in (a) are for $H_3>0$, and those seen in (b) are for $H_3<0$.

Both the spectra and the corresponding histograms of the hyperfine field distributions show high sensitivity to the amplitude and the sign of $H_3$. These features make the $^{119}$Sn Mössbauer spectroscopy a suitable tool for studying issues relevant to the SDWs which are known to be very sensitive to various kinds of lattice imperfections like foreign atoms, defects, grain boundaries, strain, etc. Consequently, in a studied sample in which any of these imperfections are present, the virgin SDWs are changed (deformed). For example magnetic impurities pin the SDWs, vanadium quenches while manganese supports the SDWs. In other words, the measurements performed on such samples do not give the proper information on the virgin SDWs.

### 2.3. SDWs and CDWs in chromium

### 2.3.1. Single crystal Cr – determination of $H_3$ and $I_2$

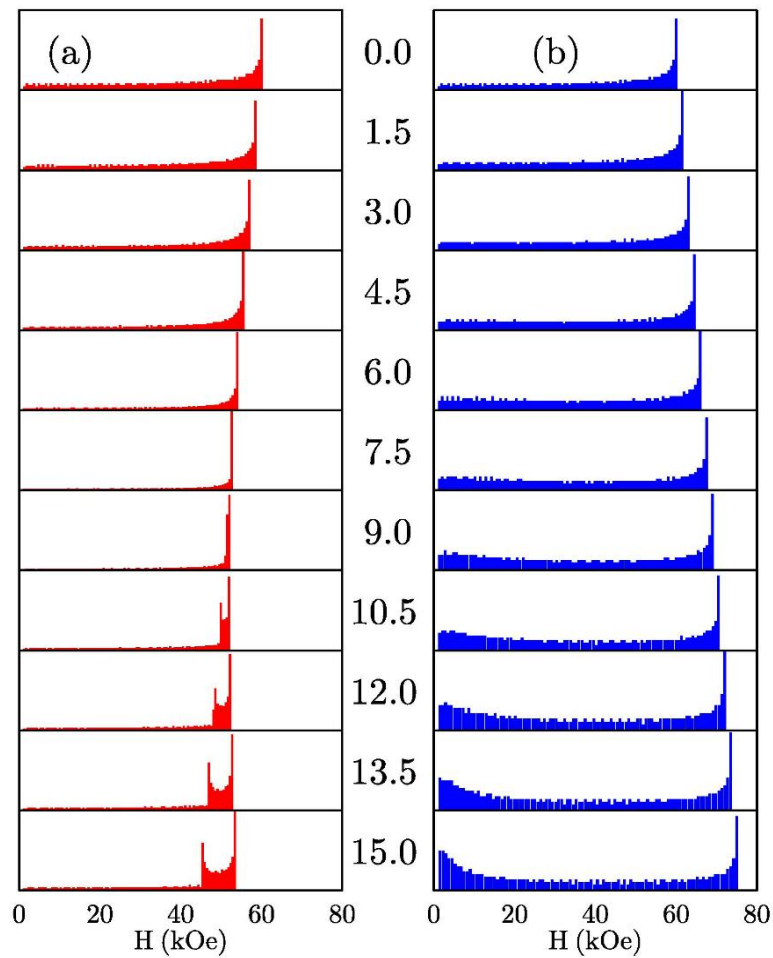

Fig. 7

Histograms of the hyperfine field distribution corresponding to the $^{119}$Sn spectra shown in Fig. 6. The labels stand for the value of the $H_3$ amplitude in kOe. The histograms shown in (a) are for $H_3>0$, and those seen in (b) are for $H_3<0$.

$^{119}$Sn Mössbauer spectra recorded at room temperature (RT) and at 4.2 K on a single-crystal sample of Cr doped with ~0.1 at% $^{119}$Sn are shown in the upper panel of Fig. 8. The shapes of the underlying SDW and CDW are shown in the bottom panel of the same figure. Concerning the SDW, one can easily notice that the amplitude increased from ~57 kOe at RT [8] to ~94 kOe at 4.2 K [18]. Furthermore, the analysis of the spectra in terms of the higher-order harmonics yielded the sign and the amplitude of $H_3$. The relative value of the latter is 2.6% at 4.2K and 1.4% at 295K. These values are in accord with 1.65% determined at 200 K by the neutron diffraction experiment [31].

The analysis of the spectra in terms of concomitant SDWs and CDWs also permitted determination of the CDW shapes and their amplitudes. Concerning the former, the shape and the periodicity is different at the two temperatures. The origin of that is unknown and the effect needs further study. On the other hand, the amplitude hardly depends on temperature and it amounts to ~5·10$^{-2}$ mm/s.

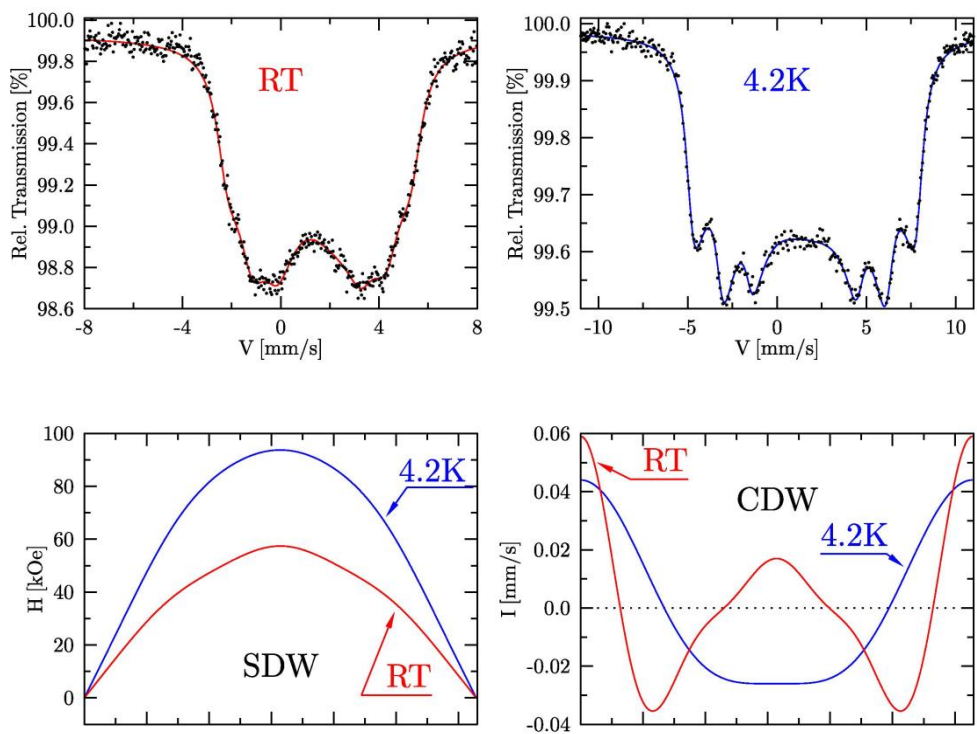

Fig. 8

$^{119}$Sn Mössbauer spectra measured on a single-crystal sample of Cr doped by diffusion with ~0.1 at% $^{119}$Sn at room temperature and at 4.2 K (upper panel), and derived therefrom shapes of the SDW (half periodicity) and CDW (full periodicity) (bottom panel) [24].

**2.3.2. Effect of grain size**

SDWs are expected to strongly interact with various kinds of lattice imperfections including grain boundaries. To verify this expectation $^{119}$Sn spectra were recorded at RT on three polycrystalline samples of Cr with different grain sizes [14]. The shape of the spectra shown in Fig. 9 evidently depends on the size of grains what can be understood in terms of interaction between the SDWs and the grain boundaries. The spectra could be successfully analyzed in terms of eq. (2) as described in detail elsewhere [14]. In addition to $H_3$, $H_5$ and $H_7$ harmonics had to be included to obtain statistically good fits.

**2.3.2. Effect of vanadium**

Substituting Cr by V drastically quenches the SDWs. Addition of ~4 at% V is enough to extinguish them completely. This quenching effect is regarded as the proof for regarding the SDWs in Cr as related to the density of electrons at the FS. $^{119}$Sn Mössbauer spectroscopy has also proved to be the pertinent method to study the issue [18]. Examples of the spectra recorded at 4.2K on single-crystals of Cr and $Cr_{100-x}V_x$ (x=0.5, 2.5, 5) doped by diffusion with a small amount of $^{119}$Sn are shown in Fig. 10. Spectral parameters related to the SDWs viz. the average hyperfine field and the maximum hyperfine field decrease with x at the same rate as the Néel temperature, the average magnetic moment and the incommensurability wave vector.

**2.3.3. Effect of surface and/or implantation**

Physical properties of surface are, in general, different than the bulk properties. Concerning magnetic properties of the surface of Cr, theoretical calculations predict that the magnetic moment of Cr atoms lying within few layers adjacent to the surface are enhanced by factor of 3.5-5 [32-35]. To verify these predictions $^{119}$Sn conversion electron spectra (CEMS) were recorded at room on a single-crystal (110) Cr foil implanted with $^{119}$Sn ions. The range of the implantation was ~16 nm what corresponds to ~2$\Lambda_a$ [21-23,25]. The measured spectrum alongside with the corresponding histogram of the hyperfine field distribution is displayed in Fig. 11. The spectrum measured in a transmission mode on a similar sample but doped with the $^{119}$Sn atoms by diffusion and the related histogram are added for the sake of comparison. The enhancement of the hyperfine field (spin-density) ranges between 2.4 and 3, so it is comparable with the theoretical predictions. The depth of the implantation corresponds, however, to about 110 monolayers i.e. is by factor 30-50 time higher than the theoretically predicted thickness. On the other hand, 110 monolayers correspond merely to ~2$\Lambda_a$, and for the SDWs it is rather the periodicity not the lattice monolayer that should be regarded as the relevant figure of merit.

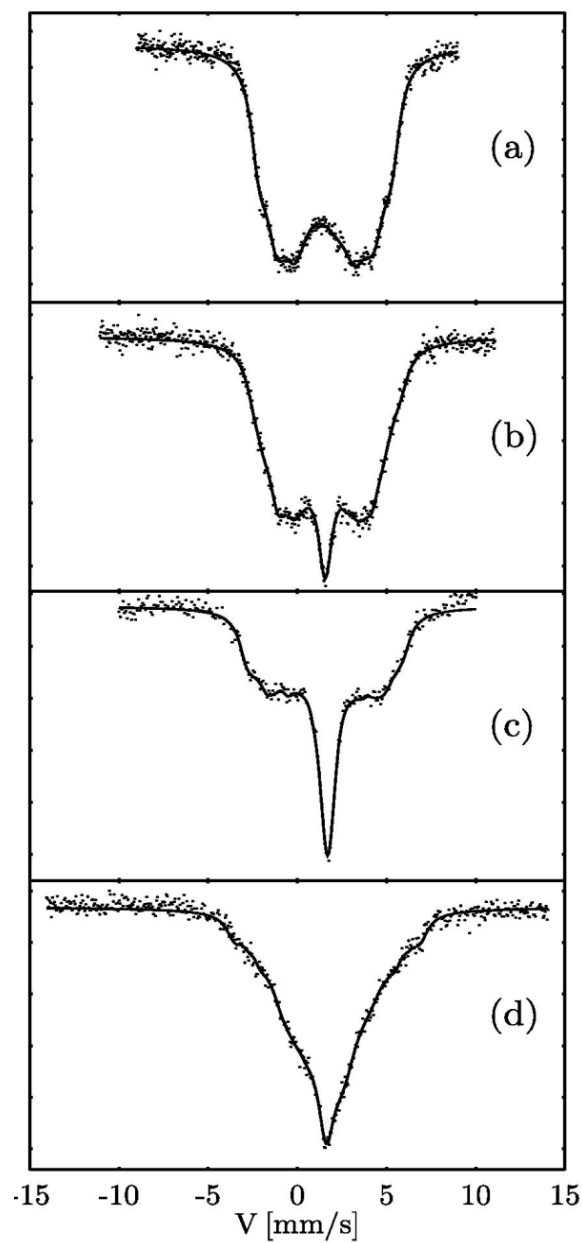

Fig. 9

$^{119}$Sn spectra recorded at RT on (a) single-crystal Cr, and (b)-(d) polycrystalline Cr with different size of grains decreasing from (b) to (d) [14].

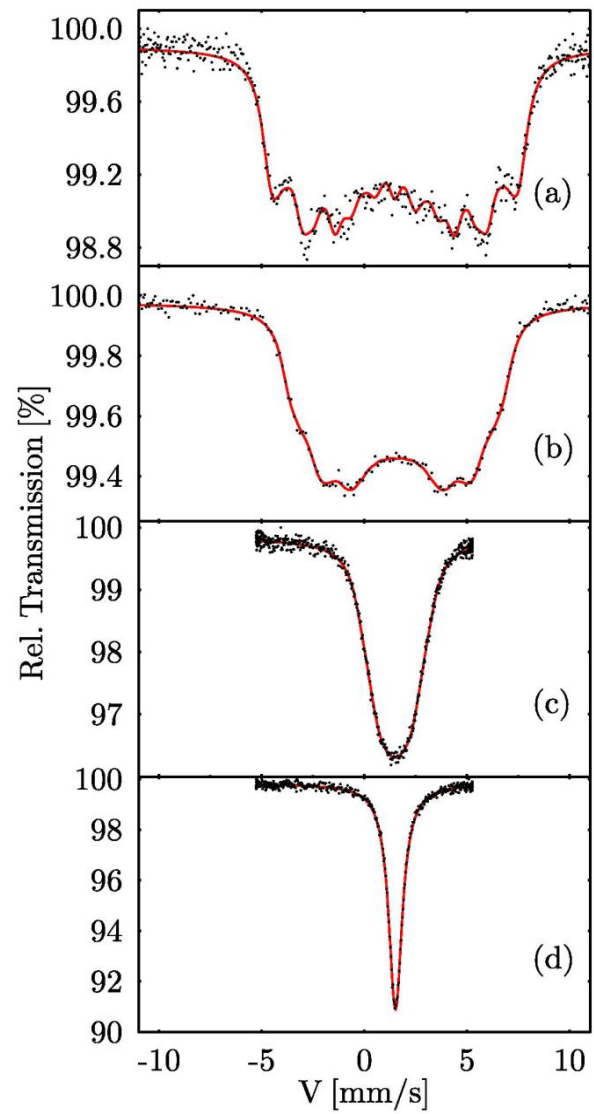

Fig. 10

$^{119}$Sn spectra recorded at 4.2K on single-crystals of (a) Cr, and (b)-(d) $Cr_{100-x}V_x$ with x=0.5 (b), x=2.5 (c) and x=5 (d) [18].

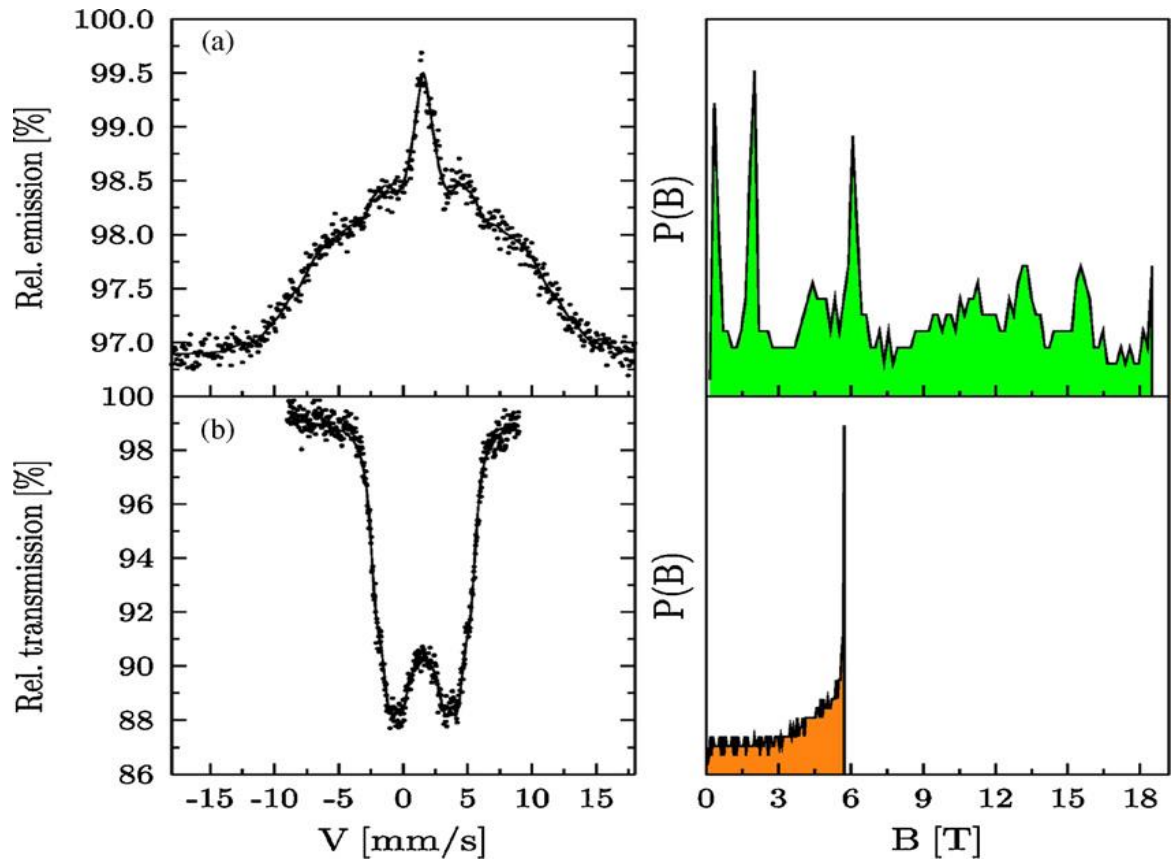

Fig. 11

$^{119}$Sn spectra recorded at RT on a single-crystal (110) Cr (a) implanted with $^{119}$Sn ions and (b) doped by diffusion with $^{119}$Sn atoms. Corresponding histograms of the hyperfine field distribution are presented, too [25].

## 3. Conclusions

The paper can be summarized with the general conclusion that the Mössbauer spectroscopy can be used as the relevant technique for investigation of virgin properties of the harmonically modulated electronic structures i.e. SDWs and CDWs. In particular it permits:

1. Distinction between the commensurate and incommensurate SDWs provided the periodicity of the former is ≤~13 lattice constants.

2. Determination of the amplitude and sign of higher order harmonics.

3. Study the effect of grain size and foreign atoms.

However,

(i) Measurements must be carried out using non-magnetic probe atoms e. g. $^{119}$Sn.

(ii) Investigated samples must be as perfect as possible i.e. single-crystals and free of impurities.